\documentclass[reqno]{amsart} % 12pt, {article} %
\usepackage{amssymb}%\usepackage{russian}\usepackage{epsf}
\usepackage{amsmath}
\usepackage{amsfonts}
\usepackage{axodraw}

\newcommand{\nc}{\newcommand}

\nc{\rnc}{\renewcommand} \nc{\beq}{\begin{equation}}
\nc{\eeq}{\end{equation}} \nc{\beqa}{\begin{eqnarray}}
\nc{\eeqa}{\end{eqnarray}}

\begin{document}

\begin{flushright} AEI-2010-148 \end{flushright}

\title[]
{\bf On the ISS model of dynamical SUSY breaking}

\author{G.~S.~Vartanov}
\address{Max-Planck-Institut f\"ur Gravitationsphysik, Albert-Einstein-Institut
14476 Golm, Germany; e-mail address: vartanov@aei.mpg.de}

\begin{abstract}
In this letter we would like to apply the superconformal index
technique to give one more evidence for the theory proposed by
Intriligator, Seiberg and Shenker (ISS) as being described by
interacting conformal field theory in its IR fixed point.

\

Keywords: Superconformal index, Dynamical Supersymmetry breaking
\end{abstract}

\maketitle

%\begin{center}
%
%%\vspace{3cm}
%
%{\bf On the ISS model of dynamical SUSY breaking}
%
%\vspace{0.3cm}
%
%
%
%{\bf G. S. Vartanov \vspace{0.5cm}\\
%{\it Max-Planck-Institut f\"ur Gravitationsphysik,
%Albert-Einstein-Institut 14476 Golm, Germany; e-mail address:
%vartanov@aei.mpg.de}}
%
%\abstract{In this Letter we would like to apply the superconformal
%index technique to give one more evidence for the theory proposed by
%Intriligator, Seiberg and Shenker (ISS) as being described by
%interacting conformal field theory in its IR fixed point.}
%
%Keywords: Superconformal index, Dynamical Supersymmetry breaking
%
%\end{center}
%
%\vspace{0.6cm}

The superconformal index technique \cite{Romelsberger1,Kinney} has
already been successfully applied in many cases for checking
conjectures of electric-magnetic duality for supersymmetric field
theories, finding new examples of dualities and also giving a lot of
new identities for mathematical functions standing behind the
superconformal indices. The Spiridonov's theory of elliptic
hypergeometric integrals is quite a new research area in mathematics
\cite{S1,S2,S3}. It was realized by Dolan and Osborn in
\cite{Dolan:2008qi} that the superconformal index is given in terms
of elliptic hypergeometric integrals and the equalities for the
superconformal indices are given by the Weyl group symmetry
transformations for elliptic hypergeometric integrals on different
root systems \cite{S2}. The first example of such integrals is given
by the elliptic beta-integral with $7$ parameters-- an extension of
the beta-integral found by Spiridonov in \cite{S1}.

Later on using the superconformal index technique in a series of
papers \cite{SV1} there were found many new Seiberg dual theories
\cite{Seiberg} (for example, supersymmetric dualities outside the
conformal window for $\mathcal{N}=1$ original Seiberg SQCD theory)
coming from the known identities for elliptic hypergeometric
integrals and vice versa a lot of new mathematical identities for
this class of special functions were conjectured. Also the same
technique showing the power of the theory of elliptic hypergeometric
integrals can be applied for extended supersymmetric theories in
order to check $S$--duality conjectures \cite{GPRR1,SV4}.

By definition the superconformal index counts the number of gauge
invariant operators which satisfy BPS condition and which cannot be
combined to form long multiplets. The $SU(2,2|1)$ space-time
symmetry group  of $\mathcal{N}=1$ superconformal algebra consists
of $J_i, \overline{J}_i$-- the generators of two $SU(2)$s which
together give the Lorentz symmetry group of a four-dimensional
theory $SO(4)$, translations $P_\mu$, special conformal
transformations $K_\mu$, $\mu=1,2,3,4$, the dilatations $H$ and also
the $U(1)_R$ generator $R$. Apart from the bosonic generators there
are supercharges $Q_{\alpha},\overline{Q}_{\dot\alpha}$ and their
superconformal partners $S_{\alpha},\overline{S}_{\dot\alpha}$. Then
taking the distinguished pair of supercharges \cite{Romelsberger1},
for example, $Q=\overline{Q}_{1 }$ and $Q^{\dag}=-{\overline
S}_{1}$, one has
\begin{equation}
\{Q,Q^{\dag}\}= 2{\mathcal H},\quad
\mathcal{H}=H-2\overline{J}_3-3R/2, \label{susy}\end{equation} and
then the superconformal index is defined by the matrix integral
\begin{eqnarray}\nonumber
&& I(p,q,f_k) = \int_{G_c} d \mu(g)\, Tr \Big( (-1)^{\rm F}
p^{\mathcal{R}/2+J_3}q^{\mathcal{R}/2-J_3}
\\ && \makebox[1em]{} \times
e^{\sum_{a} g_aG^a} e^{\sum_{k}f_kF^k}e^{-\beta {\mathcal H}}\Big),
\quad \mathcal{R}= H-R/2, \label{Ind}\end{eqnarray} where ${\rm F}$
is the fermion number operator and $d \mu(g)$ is the invariant
measure of the gauge group $G_c$. To calculate the index one should
take into account the whole space of states, but, fortunately, only
the zero modes of the $\mathcal H$ contribute to the trace because
the relation (\ref{susy}) is preserved by the operators used in
(\ref{Ind}). The chemical potentials $g_a,f_k$ are the group
parameters of the gauge and flavor symmetries and are given by
operators $G^a$ and $F^k$.

Let us consider $\mathcal{N}=1$ SYM theory based on $SU(2)$ gauge
theory with a single chiral field $Q$ in spin $I=3/2$ representation
considered by Intriligator, Seiberg and Shenker (ISS) in
\cite{Intriligator:1994rx}. The $R$--charge for the scalar component
is equal $R_Q \ = \ 3/5$. The ISS theory is conjectured to be the
simplest model of dynamical SUSY breaking (for review of the model
and further examples see \cite{Shifman:1999mv}).

In \cite{Intriligator:1994rx} two possible phases-- the confining
and the conformal-- are proposed for this theory in its IR fixed
point. If the former case is realized then deforming a theory by the
tree level superpotential $W_{tree} \ = \ \lambda Q^4$ which is a
relevant operator would dynamically break supersymmetry
\cite{Intriligator:1994rx} giving the simplest example of a theory
with dynamically broken supersymmetry. In the original paper the
authors conjectured the first possibility but later in two papers
\cite{Intriligator:2005if,Poppitz:2009kz} using different approaches
the authors give evidences for the second possibility and still the
status of this theory is open. That is why we would like to attack
this problem using the superconformal index technique. Although the
superconformal index has not yet been shown to be invariant under
the RG we rely on the latter fact being true.

The proposed dual confining phase is described by the single gauge
invariant operator coming from the electric theory which is $Q^4$
whose $R$--charge is equal to $12/5$. The only check which can be
applied to show that the original theory has confining phase is the
anomaly matching conditions. For the ISS model we have $Tr R$ and
$TrR^3$ anomalies which match in two descriptions and are given by
the following values $7/5$ and $(7/5)^3$ correspondingly.

If the ISS model has the confining phase then it can be thought as
Seiberg duality \cite{Seiberg} in the IR fixed point and according
to the R\"omelsberger's hypothesis \cite{Romelsberger1} the
superconformal indices for the ISS model and its confining phase
should coincide. And as a consequence if there is no equality
between the superconformal indices then the conformal phase takes
place. So assuming existence of the confining phase in the IR fixed
point we calculate the superconformal indices for the ISS model and
its conjectured confining phase and analyze whether two expressions
match or not. The superconformal index for the initial theory is (we
use abbreviation $I_E$ for the initial theory and $I_M$ for the
conjectured confining phase) \beq \label{1} I_E \ = \
\frac{(p;p)_\infty (q;q)_\infty}{2} \int_{\mathbb{T}}
\frac{\Gamma((pq)^{\frac{3}{10}}z^{\pm1},
(pq)^{\frac{3}{10}}z^{\pm3};p,q)}{\Gamma(z^{\pm2};p,q)} \frac{dz}{2
\pi \texttt{i} z},\eeq where $\mathbb{T}$ is the unit circle with
positive orientation and $(a;q)_\infty=\prod_{k=0}^\infty(1-aq^k)$.
Also we use the following conventions
$\Gamma(a,b;p,q):=\Gamma(a;p,q)\Gamma(b;p,q),
\Gamma(az^{\pm1};p,q):=\Gamma(az;p,q)\Gamma(az^{-1};p,q)$, where
$$
\Gamma(z;p,q)= \prod_{i,j=0}^\infty
\frac{1-z^{-1}p^{i+1}q^{j+1}}{1-zp^iq^j}, \quad |p|, |q|<1,
$$
is the so-called elliptic gamma function. For a conjectured
confining phase the superconformal index is given by one elliptic
gamma function \beq \label{2} I_M \ = \ \Gamma((pq)^{\frac
65};p,q).\eeq

Now we would like to \textit{check the conjectured equality} for the
superconformal indices coming from the assumption that there exists
a confining phase. First of all we would like to discuss the
ellipticity condition \cite{S1} for $I_E$ which states that the
ratio of the kernel taken from the integral (\ref{1}) and the same
kernel but with the substituted parameter of integration $z$ by $pz$
is the elliptic function of this parameter $z$. This condition here
from physical point of view is interpreted as having anomaly-free
gauge theory. It is satisfied trivially since the expression for the
superconformal index $I_E$, using the triplication formula for the
elliptic gamma function, \beq \Gamma(z^3;p,q) \ = \
\prod_{i,j,k=0}^2 \Gamma(z w^i p^{\frac j3} q^{\frac k3};p,q), \eeq
where $w$ is the cubic root of $1$ not equal to $1$, can be
rewritten as the higher order elliptic beta-integral with $28$
parameters in addition to $p$ and $q$ \cite{S2}.

We can calculate the series expansion in chemical potentials for
both expressions (\ref{1}) and (\ref{2}) and compare the first
several terms. Finding the series expansion in the parameter $pq$
for the superconformal index of the initial ISS theory we get \beq
I_E \ = \ 1 + \texttt{o}((pq)^{0}) \eeq and for the assumed
confining phase \beq I_M \ = \ - \frac{1}{\sqrt[5]{pq}} + 1 +
\texttt{o}((pq)^{0}).\eeq

Comparing now the first terms of series expansions we see that the
superconformal indices for two phases do not coincide showing us
that our assumption of the existing confining phase with only one
gauge invariant operator is not valid and that leads us to evidence
that the original ISS theory has interacting conformal field theory
in its IR fixed point. The same conclusion can be obtained by taking
the limit $p \rightarrow 0$ applied to the expressions for the
superconformal indices (\ref{1}) and (\ref{2}). Applying the formula
$$\lim_{p \rightarrow 0} \Gamma(z;p,q) \ = \
\frac{1}{(z;q)_\infty}$$ we see that we obtain different limits for
the $I_E$ and $I_M$.

Now we consider other well-understood models with misleading anomaly
matchings considered in \cite{Brodie:1998vv}. The theories under
consideration are $\mathcal{N}=1$ SYM with $SO$ gauge group and the
matter field $Q$ in the absolutely symmetric tensor representation
of second rank of the gauge group. The $R$--charge of the $Q$ field
is $\frac{4}{N+2}$. In \cite{Brodie:1998vv} it was shown that the
anomaly matching conditions are satisfied with the set of gauge
invariant operators $\mathcal{O}_n \ = \ \texttt{Tr} Q^n$ where
$n=2,\ldots,N$ what suggest for the confining phase with the given
above operators as fundamental fields to exist. Deforming the theory
it was shown that the confining description does not hold further so
the theory has conformal phase in its IR fixed point. We would like
to show that the same result can be obtained using the
superconformal index technique.

To calculate the superconformal indices in this case we should
distinguish between two cases, namely between even and odd rank of
the gauge group. For the case of even rank of the gauge group namely
$N=2K$ we have the following expression for the superconformal index
of the initial theory \beqa \label{SO_1} && I_E^{SO(2K)} =
\frac{(p;p)^K_\infty
(q;q)^K_\infty}{2^{K-1}K!} \Gamma^{K-1}((pq)^{\frac{2}{2K+2}};p,q) \\
\nonumber && \makebox[2em]{} \times \int_{\mathbb{T}^K} \prod_{1
\leq i < j \leq K} \frac{\Gamma((pq)^{\frac{2}{2K+2}}
z_i^{\pm1}z_j^{\pm1};p,q)}{\Gamma(z_i^{\pm1}z_j^{\pm1};p,q)}
\prod_{j=1}^K \Gamma((pq)^{\frac{2}{2K+2}} z_j^{\pm2};p,q)
\frac{dz_j}{2 \pi \texttt{i} z_j},\eeqa

The possible description of the initial theory in the IR fixed point
by the theory without gauge group and having only the gauge
invariant operators $\mathcal{O}$ bring us to the possible
expression for the index (\ref{SO_1}) by the superconformal index
which can be calculated in this confining phase \beq \label{SOb_1}
I_M^{SO(2K)} \ = \ \prod_{i=2}^{2K}
\Gamma((pq)^{\frac{i}{K+1}};p,q).\eeq

To see that the two expressions (\ref{SO_1}) and (\ref{SOb_1}) do
not coincide what is the same as there is no confining phase in the
origin of moduli space we again take the limit $p \rightarrow 0$ in
both sides and one can easily see that in the left-hand side the
limit is valid while the right-hand side does not have the valid
limit.

In the case of odd rank the superconformal index is given \beqa
\label{SO_2} && I_E^{SO(2K+1)} = \frac{(p;p)^K_\infty
(q;q)^K_\infty}{2^{K}K!} \Gamma^{K}((pq)^{\frac{2}{2K+3}};p,q) \\
\nonumber && \makebox[-2.5em]{} \times \int_{\mathbb{T}^K} \prod_{1
\leq i < j \leq K} \frac{\Gamma((pq)^{\frac{2}{2K+3}}
z_i^{\pm1}z_j^{\pm1};p,q)}{\Gamma(z_i^{\pm1}z_j^{\pm1};p,q)}
\prod_{j=1}^K \frac{\Gamma((pq)^{\frac{2}{2K+3}} z_j^{\pm1},
(pq)^{\frac{2}{2K+3}} z_j^{\pm2};p,q)}{\Gamma(z_j^{\pm1};p,q)}
\frac{dz_j}{2 \pi \texttt{i} z_j},\eeqa and if there is a possible
confining phase for this theory then the superconformal index for
this phase is given by the following expression \beq \label{SOb_2}
I_M^{SO(2K+1)} \ = \ \prod_{i=2}^{2K+1}
\Gamma((pq)^{\frac{2i}{2K+3}};p,q) \eeq

Again taking the limit $p \rightarrow 0$ we see that the analytical
properties of two sides are different which is interpreted as lack
of equality between the two expressions (\ref{SO_2}) and
(\ref{SOb_2}).

One can easily check the ellipticity condition for both cases
(\ref{SO_1}) and (\ref{SOb_1}) (for details see \cite{SV1}). Also in
both cases for even and odd ranks of the gauge group we see mismatch
of the superconformal indices of the original theory and the
conjectured confining phase. From the physical point of view this
discrepancy between the superconformal indices is realized as the
fact that in the IR fixed point the theory is given by the
interacting conformal field theory rather than by the confining
phase.

To conclude in this short Letter we describe the application of the
superconformal index technique for deriving out theories with
misleading anomaly matching \cite{Brodie:1998vv}. Also we applied
this method for the conjectured simplest model of dynamical SUSY
breaking proposed long ago in \cite{Intriligator:1994rx} and the
situation with which is not understood completely yet. Here we give
one more evidence that the proposed theory has interacting conformal
field theory description
\cite{Shifman:1999mv,Intriligator:2005if,Poppitz:2009kz} in its IR
fixed point rather than the conjectured confining theory with only
one gauge invariant operator \cite{Intriligator:1994rx}. We are
using completely different approach from the approaches in
\cite{Shifman:1999mv,Intriligator:2005if,Poppitz:2009kz} where the
analyze was based on smallness of the first coefficient of the
$\beta$ function and the possibility for calculation of the
anomalous dimensions perturbatively; the conjecture that the correct
phase in the IR fixed point of some theory is the phase with larger
conformal anomaly $a$ and the analyze of the dynamics of the theory
by its deforming on $S^1 \times \mathbb{R}^3$ correspondingly.

In \cite{SV1} it was conjectured that the anomaly matching condition
for the dual theories is given by the total ellipticity condition
for the elliptic hypergeometric integrals describing the
superconformal indices for this duality. Also it was suggested that
the ellipticity criterion is a necessary but not a sufficient
condition for a true duality. Our examples explicitly show that it
is indeed the case.

The author would like to thank Z. Komargodski for pointing out the
relevance of the observation described in this note and also thanks
F. A. H. Dolan, A. Schwimmer, M. A. Shifman and V. P. Spiridonov for
nice and stimulating discussions. The author would like to thank the
organizers of the "Simons Workshop on Mathematics and Physics
$2010$" at Stony Brook and conference "Recent Developments in
String/M-Theory and Field Theory" in Berlin for warm atmosphere
during the conferences.


\begin{thebibliography}{99}

\bibitem{Romelsberger1} C. R\"omelsberger, \textit{Counting chiral
primaries in ${\mathcal N}=1$, $d=4$ superconformal field theories},
Nucl. Phys. B {\bf 747} (2006), 329--353, arXiv:hep-th/0510060;
\textit{Calculating the superconformal index and Seiberg duality},
arXiv:0707.3702 [hep-th].

\bibitem{Kinney} J. Kinney, J. M. Maldacena, S. Minwalla and S.
Raju,  \textit{An index for 4 dimensional super conformal theories},
Commun. Math. Phys. {\bf 275} (2007), 209--254,
arXiv:hep-th/0510251.

\bibitem{S1}V. P. Spiridonov,
\textit{On the elliptic beta function}, Uspekhi Mat. Nauk {\bf 56}
(1) (2001), 181--182 (Russian Math. Surveys {\bf 56} (1) (2001),
185--186).

\bibitem{S2} V. P. Spiridonov, \textit{Theta hypergeometric
integrals}, Algebra i Analiz {\bf 15} (6) (2003), 161--215 (St.
Petersburg Math. J. {\bf 15} (6) (2003), 929--967),
arXiv:math.CA/0303205.

\bibitem{S3} V. P. Spiridonov, \textit{Essays on the theory of
elliptic hypergeometric functions}, Uspekhi Mat. Nauk {\bf 63} (3)
(2008), 3--72 (Russian Math. Surveys {\bf 63} (3) (2008), 405--472),
arXiv:0805.3135 [math.CA].

\bibitem{Dolan:2008qi}
F.~A.~Dolan and H.~Osborn, \textit{Applications of the
Superconformal Index for Protected Operators and $q$-Hypergeometric
Identities to $\mathcal{N}=1$ Dual Theories}, Nucl. Phys.  B {\bf
818} (2009), 137--178, arXiv:0801.4947 [hep-th].

\bibitem{SV1} V. P. Spiridonov and G. S. Vartanov, \textit{Superconformal
indices for ${\mathcal N}=1$ theories with multiple duals}, Nucl.
Phys. {\bf B824} (2010), 192--216, arXiv:0811.1909v4 [hep-th];
\textit{Elliptic hypergeometry of supersymmetric dualities}, Comm.
Math. Phys., to appear, arXiv:0910.5944 [hep-th];
\textit{Supersymmetric dualities beyond the conformal window},
Phys.\ Rev.\ Lett.\  {\bf 105} (2010) 061603--061606,
[arXiv:1003.6109 [hep-th]].

\bibitem{Seiberg} N. Seiberg,  \textit{Electric--magnetic duality in
supersymmetric non-Abelian gauge theories}, Nucl. Phys. {\bf B435}
(1995), 129--146, hep-th/9411149.

\bibitem{GPRR1} A. Gadde, E. Pomoni, L. Rastelli and S. S. Razamat,
\textit{ $S$-duality and $2d$ Topological QFT}, JHEP {\bf 03} (2010)
032, arXiv:0910.2225 [hep-th]; A.~Gadde, L.~Rastelli, S.~S.~Razamat
and W.~Yan, \textit{The Superconformal Index of the $E_6$ SCFT},
JHEP {\bf 1008} (2010) 107, [arXiv:1003.4244 [hep-th]].

\bibitem{SV4}
V.~P.~Spiridonov and G.~S.~Vartanov, \textit{Superconformal indices
of ${\mathcal N}=4$ SYM field theories}, arXiv:1005.4196 [hep-th].

\bibitem{Intriligator:1994rx}
K.~A.~Intriligator, N.~Seiberg and S.~H.~Shenker, \textit{Proposal
for a simple model of dynamical SUSY breaking}, Phys.\ Lett.\  B
{\bf 342} (1995) 152--154, [arXiv:hep-ph/9410203].

\bibitem{Shifman:1999mv}
M.~A.~Shifman and A.~I.~Vainshtein, \textit{Instantons versus
supersymmetry: Fifteen years later}, "ITEP Lectures in Particle
Physics and Field Theory" edited by M. Shifman, Singapore, World
Scientific, 1999. Vol. 2, pp. 485-648, arXiv:hep-th/9902018.

\bibitem{Intriligator:2005if}  K.~A.~Intriligator,
\textit{IR free or interacting? A proposed diagnostic}, Nucl.\
Phys.\  B {\bf 730} (2005) 239--251, [arXiv:hep-th/0509085].

\bibitem{Poppitz:2009kz}
E.~Poppitz and M.~Unsal, \textit{Chiral gauge dynamics and dynamical
supersymmetry breaking}, JHEP {\bf 0907} (2009) 060,
[arXiv:0905.0634 [hep-th]].

\bibitem{Brodie:1998vv}
J.~H.~Brodie, P.~L.~Cho and K.~A.~Intriligator, \textit{Misleading
anomaly matchings?}, Phys.\ Lett.\  B {\bf 429} (1998) 319--326,
[arXiv:hep-th/9802092].
\end{thebibliography}
\end{document}